\documentclass[epj]{svjour}
\usepackage{graphics}
\usepackage{cite}
\begin{document}
\title{Single-particle structure of the N~=~20, 28 isotones within the dispersive optical model }
\author{O. V. Bespalova\inst{}\thanks{\email{besp@sinp.msu.ru}} \and A. A. Klimochkina 
}                     
%
%
\institute{Skobeltsyn Institute of Nuclear Physics, Moscow State University, 119991, GSP-1, Leninskie gory 1(2), Moscow, Russia }
\date{Received: date / Revised version: date}

\authorrunning{O V Bespalova et al}
\titlerunning{Single-particle structure of the $N = 20, 28$ isotones within the dispersive optical model}

\abstract{
The neutron single-particle characteristics of the $N = 20, 28$ isotones at $8 < Z < 30$ was calculated within the dispersive optical model. The global parameters of the spin-orbit and imaginary parts of the potential as well as surface absorption independent on neutron-proton asymmetry and increased diffuseness at large neutron excess were used in the calculations.. The suitability of the global parameters to predict the evolution of the neutron single-particle structure of nuclei near the neutron drip line was investigated. The following results are in agreement with the available experimental data:  the reduction of the particle-hole energy gaps, the degeneration of the $1f_{7/2}$ and $2p$ states and then a change in the $1f_{7/2}$ , $2p_{3/2}$ level sequence and more rapid reduction of the $2p$-splitting in comparison with the $1f$-splitting with decreasing $Z$.  The predictive power of the dispersive optical model with respect to neutron-rich nuclei is demonstrated.
} 
\maketitle
\section{Introduction}
\label{intro}
Investigation of the structure of nuclei far from the $\beta$ -stability valley is one of the most intriguing questions of the modern nuclear physics. The significant progress of experimental techniques allowed discovering the specific features of exotic nuclei. The disappearance of the classic magic numbers of nucleons and the appearance of new ones are among them. The shell structure of nuclei with the sufficient neutron excess turned out to evolve essentially. The tensor forces, deformations, and the dependence of the spin-orbit interaction on the nuclear density play a significant role in the evolution of nuclear structure. 

The evolution of the neutron shell structure of the isotones with $N = 28$ manifests itself quite vividly. The low experimental excitation energy of the first $2^{+}$ level in $^{42}$Si \cite{TM12} indicate significant excitations through the particle-hole energy gap and the weakening of the classical magic number $N = 28$. The evolution of the gaps is an important feature characterizing exotic nuclei. The gap $N = 28$ is strongly suppressed in neutron rich nuclei according to the calculations within nonrelativistic and relativistic mean-field models, energy density functionals, shell model and dispersive optical model (see \cite{Sorlin08} and references therein, \cite{NP, Eb15, XWX16, UO12, SN16, Li16, Bes18}). The decrease of the gap is accompanied by a magnification of the collectivity and deformation due to the Jahn-Teller effect \cite{Ja37, RO84}, including. The deformation of isotones with $N = 28$ is supported by the tensor forces \cite {UO12, SN16}. 

The first indications of the shell closure vanishing were revealed for exotic neutron-rich nuclei with $N = 20$. In contrast to the $N = 28$ isotones, the tensor force tends to favor sphericity in the $N = 20$ isotones \cite{SN16}. The $N = 20$ isotones with $Z \leq 12$ are near or in the so called “island of inversion” \cite{W90} where the admixture of “intruder” $fp$ states is essential, and the ordering of the intruder and normal configurations is inverted. A lot of experimental and theoretical efforts were devoted to clarify the underlying physics of the “island of inversion” (see \cite{Sorlin08} and references therein, \cite{SN16, Le15, SBH08, Mac16, TS17}). The experimental observation of J$^{\pi}$ = 1$^{−}– 5^{−}$ negative-parity states gave support to the collapse of the $N = 20$ gap and the inversion between the neutron $1f_{7/2}$ and $1p_{3/2}$ levels below $Z = 12$ \cite{Le15}. 

In the present paper, the evolution of the spherical single-particle characteristics of isotones with $N = 20$ is studied in comparison with that of the $N = 28$ isotones within the dispersive optical model (DOM). The DOM was developed by C. Mahaux, R. Sartor and their coauthors (see \cite{Mah91} and references therein) at the end of the last century. In the model, the coupling of single-particle motion with more complex configurations is taken into account by the dispersive component of the dispersive optical-model potential (DOMP) that leads to an increase in the concentration of the single-particle levels near the Fermi energy $E_{F}$ as compared to calculations in the Hartree-Fock approximation. Very good agreement of the calculated nucleon scattering cross sections and single-particle properties of the stable nuclei with the experimental data was achieved for nuclei at mass number $40 \leq A \leq 208$ and energy  $ 70 \leq E \leq 200$ MeV using DOMP \cite{Mah91, Mul11, BR15, Dick17, Cap05 }. The DOM permits also to trace the evolution of the nuclear single-particle structure when $N$ and $Z$ numbers change in a wide region \cite{Mul11, BR15}. In this study, the predictive power of DOM \cite{Mah91} when choosing the global parameters, the diffuseness parameter of the neutron potential and the dependence of the surface absorption on neutron excess is verified with respect to the $N = 20, 28$ isotones near the neutron drip line. 

\section{The model and its parameters}
\label{sec:2}

The DOM mean field is unified for negative and positive energies so that the single-particle properties of the nucleon bound states and elastic nucleon scattering cross sections can be calculated by the unified manner. The neutron-nucleus spherical optical model potential is given as:
\begin{eqnarray*}
U(r,E)=U_{p}(r,E)-U_{so}(r,E),
\end{eqnarray*}
\noindent
 where $U_{p}(r,E)$ is the central potential, $U_{so}(r,E)$ is the spin-orbit potential. The central real part of the DOMP is represented by the sum of three terms, namely, the Hartree-Fock type potential $V_{HF}$ and the volume, $ \Delta V_{s}$, and surface, $\Delta V_{d}$, dispersive components, so  is:
\begin{eqnarray*}
U_{p}(r,E)=V_{HF}(E)f(r,r_{HF},a_{HF})+\\
+\Delta V_{s}(E)f(r,r_{s},a_{s})-4a_{d}\Delta V_{d}(E)\frac{d}{dr}f(r,r_{d},a_{d})+\\
+iW_{s}(E)f(r,r_{s},a_{s})-4ia_{d}W_{d}(E)\frac{d}{dr}f(r,r_{d},a_{d}) ,    
\end{eqnarray*}
where $f(r,r_i, a_i)$ is the Saxon-Woods function. The dispersive components are determined from the dispersion relation:
\begin{eqnarray*}
\Delta V_{s(d)}(E)=(E_{F}-E)\frac{P}{\pi}\int\limits_{-\infty}^{\infty}\frac{W_{s(d)}(E')}{(E'-E_F)(E-E')}dE' .
\end{eqnarray*}
(P denotes the principal value). The components      $\Delta V_{s,d} (r, E)$ effectively take into account the nucleon volume $(s)$ and surface $(d)$ correlations. The energy dependences of the surface $W_d$ and volume $W_s$ components of the imaginary part of DOMP at $ E > E_F$ were chosen to be of the form:
$$W_{s}(E)=w_1\frac{(E-E_{p})^2}{(E-E_p)^2 +w_2^2} ,$$
$$W_{d}(E)=d_{1}\frac{(E-E_{p})^2 \exp[-d_2(E-E_p)]}{(E-E_p)^2 +d_3^2} .$$
The components $W_{s,d} (E)$, are assumed to be symmetric relative to the Fermi energy and equal to zero in the range $|E-E_F|<|E_p-E_F|$. The energy $E_F$ was determined from the AME16 data \cite{Wang17} on the neutron separation energy  from nucleus with the numbers $(N, Z)$ and $(N + 1, Z)$:
\begin{eqnarray*}
E_{F}=-\frac{[S_{n}(N,Z) + S_{n}(N+1, Z)]}{2} .
\end{eqnarray*}
The expression for estimating the energy $E_p$ was proposed in \cite{Mul11}:
\begin{eqnarray*}
|E_{F} - E_{p}|=0.8 [\frac{\Delta S_{n}}{2} + min(\Delta S_{n}, \Delta S_{p})] ,
\end{eqnarray*}
where $$\Delta S_{i} = (S_{i}(A) - S_{i}(A+1)),  i = n, p. $$
The $E_p$ values which have been determined in this paper are presented in table~\ref{tab:Param}.
The energy dependence of the depth parameter $V_{HF}$ was parameterized by the exponential function:

\begin{equation}
V_{HF}(E)=V_{HF}(E_F)\exp\left({-\frac{\gamma (E-E_F)}{V_{HF}(E_F)}}\right) .
\label{Vhf}
\end{equation}

Parameterization (\ref{Vhf}) was used to calculate the bound state properties for all nuclei with the exception of $^{40,48}$Ca for which experimental scattering data are available in a wide energy range. In these nuclei, the dependence was parameterized by an expression that allows ones to describe the scattering data as well as the bound states data:

\begin{eqnarray*}
V_{HF} (E) = V_{HF}(E_{F}) - \lambda (E - E_{F}) ,    \mbox{ } E_{nlj} \leq E_F, 
\end{eqnarray*}
\begin{eqnarray*}
V_{HF}(E) = V_{HF}^{1}(E_{F})+ V_{HF}^{2}(E_{F}) \times \\
\times \exp\left({\frac{-\lambda(E-E_{F})}{V_{HF}^{2}(E_{F})}}\right), \mbox{ } E_{nlj} > E_F, 
\end{eqnarray*}
\begin{equation}
V_{HF}(E_{F})=V_{HF}^{1}(E_{F})+V_{HF}^2(E_{F}).
\end{equation}  \label{Vhfpods}
Single-particle energies were calculated by solving the Schrödinger equation with the real part of the DOMP:
\begin{eqnarray*}
\left[\frac{-\nabla^{2}}{2m}+V(r,E_{nlj}) \right]\Phi _{nlj}(\vec{r}) = E_{nlj}\Phi _{nlj}(\vec{r}) .
\end{eqnarray*}
Then the radial part  $u_{nlj}(r)$ of the wave function $\Phi_{nlj}(\vec{r})$  was corrected to take into account the nonlocality effect:
\begin{eqnarray*}
\bar{u}_{nlj}=C_{nlj}(m^{*}_{HF}(r,E)/m)^{1/2}u_{nlj}(r) .
\end{eqnarray*}
The ratio of the Hartree-Fock effective mass of nucleon $m_{HF}^{*}(r,E)$ to its total mass m is given by the expression:
\begin{eqnarray*}
m^{*}_{HF}(r,E)/m=1-\frac{d}{dE}V_{HF}(r,E) .
\end{eqnarray*}
The coefficient $C_{nlj}$ was determined by normalizing $\bar{u}_{nlj}(r)$ to unity. The root-mean-square radii of the bound neutron states were calculated as following: 
\begin{equation}
R_{nlj}^{rms} = \left [ \int_{0}^{\infty } \bar{u}^2_{nlj}(r)r^2dr\right ]^{1/2} .
\label{Rrms}
\end{equation}

\begin{figure}
\resizebox{6,5 cm}{!}{%
\includegraphics{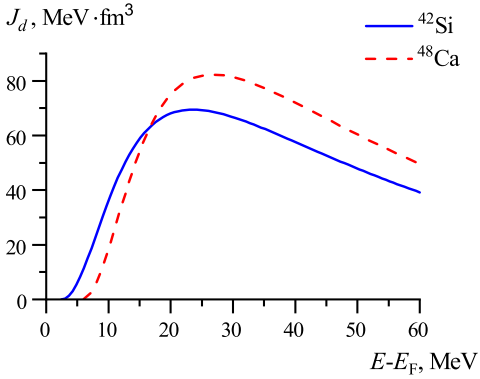}
}
\caption{Volume integral $J_d$ of the surface imaginary part of the DOMP for $^{48}$Ca and $^{42}$Si (taken from [9]).}
\label{fig:Jd}      
\end{figure}

\begin{table}
\caption{The parameters of $V_{HF}(E_F)$ (in MeV) and $E_p$ (in MeV) of neutron DOMP for the $ N =$ 20, 28 isotones.}
\label{tab:Param}       
\begin{center}
\begin{tabular}{ccccc}
\hline\noalign{\smallskip}
&\multicolumn{2}{c}{\small{$N=$ 20}} & \multicolumn{2}{c}{\small{$N=$ 28}}\\
\noalign{\smallskip}\hline\noalign{\smallskip}
Nucleus & $ V_{HF}(E_F)$ & $ -E_p $ & $ V_{HF}(E_F) $ & $ -E_p $ \\
\noalign{\smallskip}\hline\noalign{\smallskip}
O&46.5&0.13& & \\
Ne & 51.5 & 1.9 & 42.9 & -1.3 \\
Mg & 52.5 & 4.0 & 45.0 & -3.8 \\
Si & 51.9 & 3.5 & 49.2 & 0.06 \\
S & 52.5 & 4.0 & 50.2 & 1.3 \\
Ar & 55.2 & 7.0 & 52.0 & 0.4 \\
Ca & 58.0* & 2.0 & 53.4* & 1.8 \\
Ti & 60.9 & 14.0 & 54.0 & 3.6 \\
Cr & 61.7 & 15.0 & 55.0 & 5.2 \\
Fe & 62.9 & 18.6 & 56.1 & 6.7 \\
Ni & 64.0 & 16.0 & 57.7 & 5.8 \\
Zn &  &  & 59.4 & 10.8 \\
Ge &  &  & 60.3 & 11.8 \\
\hline\noalign{\smallskip}
\multicolumn{5}{p{8cm}}{\small{$^{*)}$ $V_{HF}(E_F) = V_{HF}^2(E_F), V_{HF}^1(E_F)= 0 $ }}\\
\end{tabular}
\end{center}
\end{table}
First of all, the global parameters KD \cite{KD} of the traditional (nondispersive) optical-model potential for the volume $W_s$ and surface $W_d$ imaginary part of as well as spin-orbit interaction were used to construct the neutron DOMP and to trace the evolution of the single-particle properties of the $N = 20, 28$ isotones. As an example, the energy dependences of the volume integral  $J_d(E)$ of the surface imaginary potential $W_d(r,E)$, which were calculated with the global parameters KD and $E_p$ from table~\ref{tab:Param}, are shown in fig.~\ref{fig:Jd} for $^{48}$Ca and $^{42}$Si isotones with $N = 28$. The energy range, where $J_d(E)$  is equal to 0, is substantially less for the neutron-rich $^{42}$Si isotone than for the stable $^{48}$Ca isotone, so that the surface absorption of neutrons by the $^{42}$Si isotone is stronger than by the $^{48}$Ca isotone near the Fermi energy in spite of the sufficient neutron excess in $^{42}$Si nucleus.

The strength parameter $V_{HF}(E_F)$  (see table~\ref{tab:Param}) was determined from the fitting to the Fermi energy. Geometrical parameters $r_{HF}$ was chosen to be equal to $r_V$ \cite{KD}. The parameter $\gamma = 0.410~(\ref{Vhf}) $ was determined by agreement with the experimental energy $E_{nlj}^{exp}$  \cite{VV90} of $1s_{1/2}$ state for $^{40}$Ca and then was fixed for all nuclei. Then the expressions (\ref{Vhfpods}) were used for $^{40}$Ca and $^{48}$Ca and the parameters $\lambda = $0.570 and 0.580 were determined respectively from the fit to the experimental elastic scattering cross sections \cite{Mul11, Tor82, Hon86, Os04, Hic90}. \\
It was obtained \cite{Mul11} that the surface absorption $W_d$ of neutrons depends on neutron-proton asymmetry weaker than the surface absorption of protons for nuclei with $N > Z$. The imaginary part $W_d$ of the global potential KD depends on $(N - Z)/A$ in the same way for neutrons and protons (with different signs). The calculation with the surface absorption $W_d$, which is independent on neutron-proton asymmetry, was additionally carried out to study how its dependence on $(N - Z)/A$ affects the neutron single-particle properties of the $N = $20 isotones with a sufficient neutron excess.

In the DOM \cite{Mah91}, the occupation probabilities $N_{nlj}^{DOM}$  are completely determined by the energy dependence of the imaginary part of the DOMP and are calculated using the approximated formulae. At present, it is not clear enough how the surface imaginary potential depends on neutron-proton asymmetry at sufficient neutron excess. This introduces uncertainty in $N_{nlj}^{DOM}$ values. In addition, the occupation probabilities $N_{nlj}^{DOM}$ often correspond to the total number of $N (Z)$ of a nucleus with an error of more than 1 nucleon. In the present paper, occupation probabilities $N_{nlj}$  were evaluated by the expression of the BCS theory:
\begin{equation}
N_{nlj} = 1/2 \cdot\left ( 1-\frac{(E_{nlj} - E_F)}{\sqrt{(E_{nlj}-E_F)^2+(\Delta)^2}} \right ) .
\label{Nnlj} 
\end{equation}
with the energies $E_{nlj}$  calculated by DOM. The pairing gap parameter $\Delta $  was taken as:
\begin{eqnarray*}
\Delta =1/4\left \{ S_n(A+1) - 2S_n(A)+S_n(A-1) \right \}  .
\end{eqnarray*}
The separation energies $S_n$ from the HFB-24 mass formula \cite{Goriely13}, were used to calculate $\Delta$  when AME16 data were missing. The agreement of the total number of neutrons $N_{nlj}=\sum (2j+1)N_{nlj}$  with the number $N$ of nucleus is achieved within an accuracy of $< 1$ neutron if the occupation probabilities (\ref {Nnlj}) are used.

\section{Calculation results }
\label{sec:3}

\begin{figure}
\resizebox{8,5 cm}{!}{%
\includegraphics{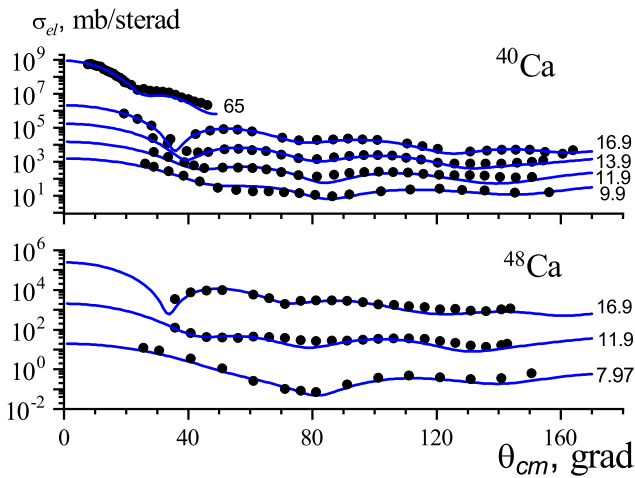}
}
\caption{ Neutron differential elastic scattering cross sections $\sigma_{el}(\theta)$  for $^{40,48}$Ca. Circles are the experimental data \cite{Mul11, Tor82, Hon86, Os04, Hic90}, lines are the calculation with the DOMP. The numbers near the graphs are the neutron energies in MeV.}
\label{fig:SigmaEl}      
\end{figure}
\begin{figure}
\resizebox{8 cm}{!}{%
\includegraphics{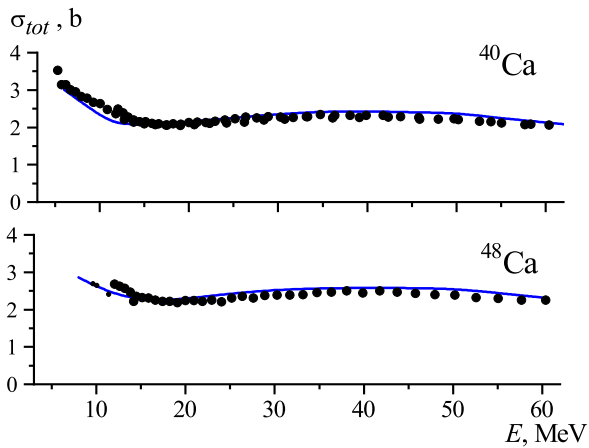}
}
\caption{Neutron total interaction cross sections $\sigma_{tot}$  for $^{40,48}$Ca. Circles are the experimental data \cite{Har85, Sh10, Fin93}, lines are the calculation with the DOMP.}
\label{fig:SigmaTOT}      
\end{figure}

The calculated neutron differential elastic scattering cross sections $\sigma_{el}(\theta)$  and the total neutron interaction cross sections $\sigma_{tot}$  at $E < $65 MeV are shown in fig.~\ref{fig:SigmaEl} and ~\ref{fig:SigmaTOT}, respectively, compared with the experimental data \cite{Mul11, Tor82, Hon86, Os04, Hic90, Har85, Sh10, Fin93} for the $^{40, 48}$Ca. The initial version of DOM \cite{Mah91} allowed us to achieve a good description of the elastic scattering data with the minimum number of fitted parameters. The calculated total interaction cross sections  $\sigma_{tot}$ slightly overestimate the experimental data in the range of medium energies and underestimate the data in the low-energy range. A better description of $\sigma_{tot}$ was achieved by the original nonlocal dispersive optical model with a greater number of parameters \cite{Mul11}. 

The calculated single-particle energies  $E_{nlj}$ and the occupation probabilities $N_{nlj}$ for neutron bound states in $N= $20 and $N= $28 isotopes are shown in table~\ref{tab:En} compared with the experimental (or evaluated) data \cite{Bob10, Besp18, Gau06, Graw01}. The data \cite{Bob10} were determined by the joint evaluation \cite{Bob89} of the stripping and pickup reaction data. Good agreement was achieved using the global parameters KD of the imaginary and spin-orbit parts of DOMP. Then, the global parameters KD were applied to predict the evolution of the single-particle structure of neutron-rich $N = $20, 28 isotones.

 \begin{figure} 
\resizebox{8,5 cm}{!}{%
\includegraphics{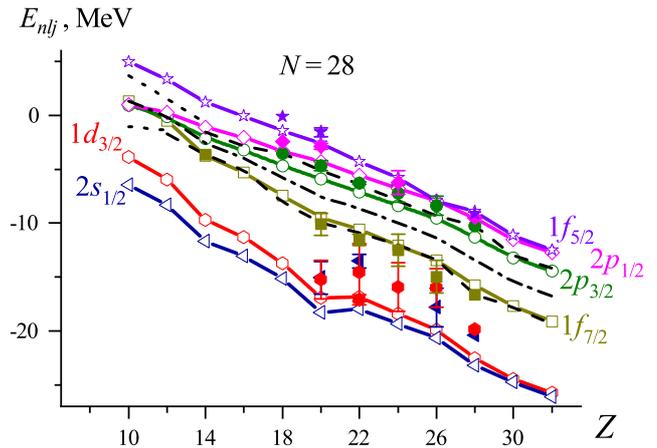}
}
\caption{ Neutron single-particle energies $E_{nlj}$ for the N = 28 isotones. Solid symbols are the experimental data \cite{Bob10, Besp18, Gau06, Graw01}, solid lines with open symbols are the calculation with the DOMP \cite{Bes18}, dash-dotted line is the energy $E_F$, dashed and dotted lines are the neutron separation energies $-S_n(N=28, Z)$ and $ -S_n(N=29, Z)$  (for even Z) \cite{Wang17} and \cite{Goriely13}, respectively.}
\label{fig:Enlj28}      
\end{figure}




Argon with $Z = $ 18 is the lowest even-$Z$ nucleus exhibiting the $N =$ 28 closed shell according to the experimental data \cite{Mei15}. At further $Z$ decrease, excitations through the $N =$ 28 gap intensify so that $^{42}$Si is characterized by the low experimental excitation energy $E(2^+_1) = $ 742(8) keV \cite{TM12}. The evolution of the $N =$ 28 shell gap were intensively investigated theoretically (see \cite{Sorlin08} and references therein, \cite{NP, Eb15, XWX16, UO12, SN16, Li16, Bes18}) and quenching of the $N = $28 gap was predicted confidently. Evolution of the single-particle spectra of the $N =$ 28 isotones, calculated by the DOM, is shown in fig.~\ref{fig:Enlj28} in comparison with the experimental and evaluated data \cite{Bob10, Besp18}. The main feature of the evolution is a sharp decrease of the gap between the $1f_{7/2}$ and $2p_{3/2}$ levels with an increase in the neutron excess at $Z <$ 18. The calculated gap reduced by 3.4 MeV from $^{48}$Ca to $^{42}$Si. This value is comparable with the difference $\left [ \left \{ S_n(^{48}Ca) -S_n(^{49}Ca) \right \} -\left \{ S_n(^{42}Si)-S_n(^{43}Si) \right \}\right ] =$ 2.7 MeV \cite{Wang17} and is larger than the estimation of 1 MeV, obtained by assuming that the gap evenly reduces by 330 keV per pair of remote protons, as it was measured for $^{47}$Ar in comparison with $^{49}$Ca \cite{Gau06}. Reduction of the gap between $^{48}$Ca and $^{44}$S also agrees with the calculations in the framework of relativistic continuum Hartree-Bogoliubov theory \cite{XWX16}. The gap disappears at $Z =$ 12 so that $1f_{7/2}$, $2p_{3/2} $ and $2p_{1/2}$ states degenerate. It can lead to deformation of the nucleus as a result of the Jahn-Teller effect \cite{Ja37, RO84} caused by a particular coherent superposition of the (near) degenerated single-particle states. It was shown \cite{UO12, SN16} that the Jahn-Teller mechanism of deformation is supported by the tensor interaction and plays a significant role in the appearance of the oblate shape of $^{42}$Si. 
\begin{table*}
\caption{Neutron single-particle energies  $E_{nlj}$ (MeV) and occupation probabilities $N_{nlj}$  for $N=$ 20, 28 isotones }
\label{tab:En}       
\begin{center}
 \begin{tabular}[h]{p{1.6cm}|p{1.6cm}p{1.6cm}p{1.6cm}p{1.6cm}|p{1.8cm}p{1.6cm}p{1.6cm}p{1.3cm}}
\hline\noalign{\smallskip}
subshell &$-E_{nlj}$&$-E_{nlj}$&$N_{nlj}$&$N_{nlj}$&$-E_{nlj}$&$-E_{nlj}$&$N_{nlj}$&$N_{nlj}$  \\
&exp&DOM&exp&(\ref{Nnlj})&exp&DOM&exp&$(\ref{Nnlj})$  \\
\noalign{\smallskip}\hline\noalign{\smallskip}
 & \multicolumn{4}{|c|}{$^{34}$Si} & \multicolumn{4}{c}{$^{40}$Ca}  \\
\noalign{\smallskip}\hline\noalign{\smallskip}
$1f_{5/2}$& &-2.60& &0.02 &1.46(20)&1.21&0.02(2)&0.01\\
$2p_{1/2}$& &0.37& &0.04&4.27(43)&3.26&0.01(1)&0.02\\
$2p_{3/2}$&1.57(4)&1.26& &0.06&6.10(67)&4.74&0.02(2)&0.03\\
$1f_{7/2}$&2.47(4)&2.09& &0.09&7.52(75)&7.86&0.02(2)&0.09\\
$1d_{3/2}$&7.51(2)&7.10& &0.86&15.2(20)&15.35&0.83(3)&0.89\\
$2s_{1/2}$&8.52(2)&9.24& &0.95&17.2(14)&16.95&$>$0.92&0.94\\
\noalign{\smallskip}\hline\noalign{\smallskip}
 & \multicolumn{4}{|c|}{$^{46}$Ar} & \multicolumn{4}{c}{$^{48}$Ca}  \\
\noalign{\smallskip}\hline\noalign{\smallskip}
$1f_{5/2}$&0.1&1.45& &0.03&1.57(40)&2.58&0.03(3)&0.03\\
$2p_{1/2}$&2.42&3.29& &0.08&2.87(28)&4.29&0.00(0)&0.07\\
$2p_{3/2}$&3.55&4.69& &0.22&4.68(47)&5.90&0.01(1)&0.17\\
$1f_{7/2}$& &7.45& &0.82&10.1(10)&9.49&1.00(0)&0.86\\
$1d_{3/2}$&&13.75& &0.99&15.2(18)&16.90&0.99(1)&0.99\\
$2s_{1/2}$& &15.17& &0.99&15.1(15)&18.32&1.00(0)&0.99\\
\noalign{\smallskip}\hline\noalign{\smallskip}
 & \multicolumn{4}{|c|}{$^{50}$Ti} & \multicolumn{4}{c}{$^{52}$Cr}  \\
\noalign{\smallskip}\hline\noalign{\smallskip}
$1f_{5/2}$&4.14(67)&4.32&0.13(4)&0.04& &5.78& &0.04\\
$2p_{1/2}$&4.60(24)&5.58&0.03(3)&0.07&6.24(110)&6.79&0.08(6)&0.06\\
$2p_{3/2}$&6.36(7)&7.11&0.09(1)&0.18&7.23(80)&8.38&0.05(3)&0.16\\
$1f_{7/2}$&10.9(6)&10.56&0.89(4) &0.86 &12.5(15)&12.07&0.92(7)&0.89\\
$1d_{3/2}$&14.60(27)&16.88&0.87(14)&0.99&15.9(23)&18.40& & \\
$2s_{1/2}$& &17.97&0.84(2)&0.99& &19.35& & \\
\noalign{\smallskip}\hline\noalign{\smallskip}
 & \multicolumn{4}{|c|}{$^{54}$Fe} & \multicolumn{4}{c}{$^{56}$Ni}  \\
\noalign{\smallskip}\hline\noalign{\smallskip}
$1f_{5/2}$& &7.93& &0.04&9.14&8.92& &0.05\\
$2p_{1/2}$& &7.93& &0.05&9.48&9.54& &0.06\\
$2p_{3/2}$&8.4(9)&9.54&0.07(3)&0.14&10.25&11.3& &0.15\\
$1f_{7/2}$&14.9(15)&13.45&0.95(2)&0.89&16.65&15.77& &0.86\\
$1d_{3/2}$&16.0(18)&19.87&0.80(6)& &19.84&22.55& &0.99\\
$2s_{1/2}$&17.8(18)&20.67& & &20.40&23.18& &0.99\\
\noalign{\smallskip}\hline
\end{tabular}
\end{center}
\end{table*}
\begin{figure}
\resizebox{8 cm}{!}{%
\includegraphics{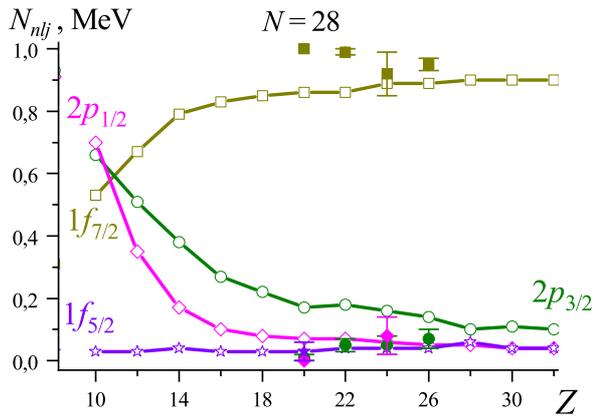}
}
\caption{The evaluated neutron occupation probabilities $N_{nlj}$  for the $N =$ 28 isotones. Solid symbols are the experimental data \cite{Bob10}, lines with open symbols are $N_{nlj}$ (\ref{Nnlj}) (was taken from \cite{Bes18}. }
\label{fig:Nnlj28}      
\end{figure}

Fig.~\ref{fig:Enlj28} demonstrates that the energy of the last predominantly occupied $1f_{7/2}$ state is close to the neutron separation energy (with the opposite sign) $ - S_n(N = 28, Z )$ \cite{Wang17} from even-even nuclei down to $Z =$ 14, and the energy of the first predominantly unoccupied $2p_{3/2}$ state is close to the neutron separation energy $-S_n(N = 29, Z)$ \cite{Wang17} down to $Z = $20. At further $Z$ decrease, the spin-orbital splitting of $2p$ states becomes rather small, and the averaged energy of $2p$ states is close to the energy $- S_n(N = 29, Z)$ down to $Z =$ 14. At $Z < $14, such a pattern changes dramatically and the correspondence between the energies of $1f_{1/2}$, $2p$ states and $-S_n(N, Z)$, $-S_n(N+1, Z)$ \cite{Goriely13} is no longer observed indicating the complete disappearance of the $N =$ 28 magic number.

The evaluated occupation probabilities   $N_{nlj}$ (\ref{Nnlj}) of the neutron states near the Fermi energy of $N =$ 28 isotones are shown in fig.~\ref{fig:Nnlj28} in comparison with the experimental data \cite{Bob10}. The probabilities  $N_{nlj}$ of $2p$ and $1f_{7/2}$ states increase and decrease accordingly with increasing neutron excess so that   $N_{2p} > N_{1f_{7/2}}$   at $Z < $12. 

The island of inversion at $N =$ 20 has been the subject of intense theoretical works (see \cite{Sorlin08} and references therein, \cite{SBH08, Mac16, TS17}). The shell-model calculations with the effective nucleon-nucleon interaction obtained from the extended Kuo-Krenciglowa theory \cite{TS17} took into account the tensor and three-nucleon forces (3NF). The results predict that the gap N = 20 disappears near Z = 8, while the tensor and the 3NF forces led to an increase in the $N =$ 20 gap and the energies of $sd-pf$ orbits, respectively. The change of the $2p_{3/2} - 1f_{7/2}$ ordering for the $N = $20 isotones at $Z$ near 10 was obtained with the monopole proton-neutron interactions in the Monte Carlo shell-model approach \cite{Uts99}. The inversion between the $f_{7/2}$ and $p_{3/2}$ levels was supported experimentally by the presence of a multiplet of $J^{\pi} = 1^{-} - 5^{-}$ negative-parity states observed under spectroscopic study of $^{28}$Na \cite{Le15}. 

\begin{figure}
\resizebox{8,5 cm}{!}{%
\includegraphics{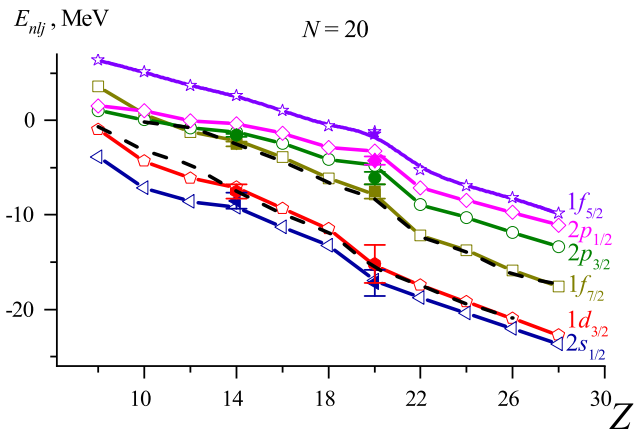}
}
\caption{Neutron single-particle energies $E_{nlj}$  for the $N =$ 20 isotones. Solid symbols are the experimental and evaluated data \cite{Bob10, Besp18}, solid lines with open symbols are the calculation with the DOMP, dashed and dash-dotted lines are the same as in fig.~\ref{fig:Enlj28}.}
\label{fig:Enlj20}      
\end{figure}

The change of the $2p$ and $1f_{7/2}$ level sequence occurs inside of the neutron drip line at $N =$ 20 in contrast to $N =$ 28 (see fig.~\ref{fig:Enlj20}). According to the DOM, the energy of the $2p_{3/2}$ state is very close to the energy of the $1f_{7/2}$ state at $Z =$ 12, so that the former becomes the first predominantly unoccupied state at further $Z$ decrease in $N =$ 20 isotones. So, the gap $N =$ 20 between the $1d_{3/2}$ state below the $E_F$ and the $2p_{3/2}$ state above the $E_F$ is formed. The number $(2j+1)$ is equal to 4 for the $1d_{3/2}$ and $2p_{3/2}$ states. It corresponds to a significant increase in the contribution of $4p4h$ excitations to the ground state of $^{32}$Mg which was predicted by shell model calculations \cite{Mac16, TS17}. This contribution determines the position of this nucleus on the shore of the island of inversion. 
 
\begin{figure}
\resizebox{8,2 cm}{!}{%
\includegraphics{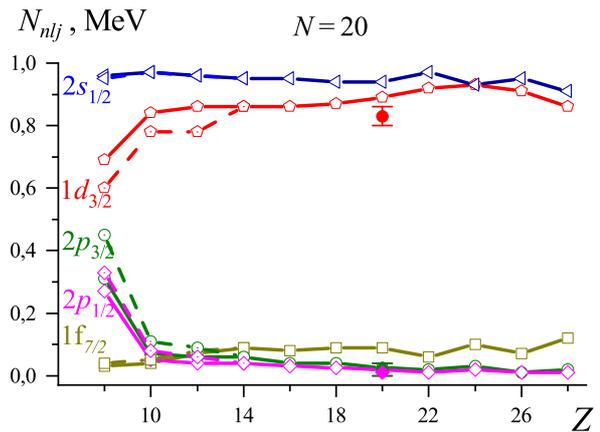}
}
\caption{The evaluated neutron occupation probabilities $N_{nlj}$  for the $N =$ 20 isotones. Solid symbols are the experimental data \cite{Bob10}, solid lines with open symbols are $N_{nlj}$ (\ref{Nnlj}) corresponding to the energies $E_{nlj}$  calculated with the global parameters KD of the surface absorption $W_d$, dashed lines are the calculation with $W_d$, which is independent of $(N - Z)/A$.}
\label{fig:Nnlj20}      
\end{figure}

The $N =$ 20 gap reduces with $Z$ decreasing due to the largest slope of the dependence  $E_{nlj}(Z)$ for the $1d_{3/2}$ state. This is the result of the strong $d_{5/2} - d_{3/2}$ proton-neutron monopole interaction when the proton $d_{5/2}$ is filled according to \cite{Ots01}. DOM does not explicitly account for tensor forces. But contribution of correlations to the evolution of single-paritcle characteristics is described by the dispersive components of DOMP. The DOM mean field lead to the quenching of the $N =$ 20 gap accompanied by the widening of the $N =$ 16 gap between the $2s_{1/2}$ and $1d_{3/2}$ states. The appearance of the $N =$ 16 shell gap was predicted earlier \cite{Oz00, Ots01}. The corresponding evaluated occupation probabilities $N_{nlj}$  (\ref{Nnlj}) of the $2p$ states increase with the neutron excess and exceed $N_{nlj}$  of the $1f_{7/2}$ state near $Z =$ 12 while $N_{nlj}$  of the $1d_{3/2}$ state decrease (see fig.~\ref{fig:Nnlj20}).

\begin{figure}
\resizebox{8,5 cm}{!}{%
\includegraphics{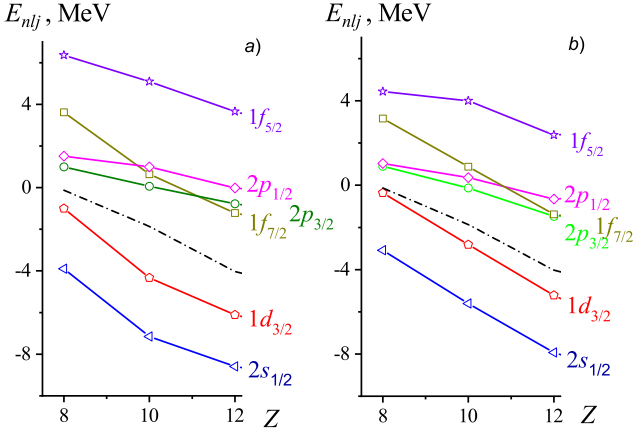}
}
\caption{Neutron single-particle energies $E_{nlj}$  for the $N =$ 20 isotones with the sufficient neutron excess. Lines are the calculation with the global parameters KD of the surface absorption $W_d~(a)$ and with $W_d$ independent of $(N - Z)/A~(b)$.}
\label{fig:Enlj20Big}      
\end{figure}

The surface absorption substantially influences on the single-particle energies near the Fermi energy through the dispersive components. The surface absorption of neutrons was obtained \cite{Mul11} to be weaker dependent on neutron-proton asymmetry in comparison with proton ones for stable nuclei with $Z < N$. To clarify the influence of this dependence on the neutron single-particle properties of nuclei with the neutron excess, we also calculated the energies $E_{nlj}$ and the occupation probabilities $N_{nlj}$~(\ref{Nnlj}) for the $N =$ 20 isotones with $Z <$ 14 assuming that $W_d$ does not depend on $(N - Z)/A$. The resulting $N =$ 20 gap decreases more pronounced than the gap calculated with the global parameters KD (see fig.~\ref{fig:Enlj20Big}). Besides, a change in the $1f_{7/2} - 2p_{3/2}$ ordering occurs already at $Z = $12, in the negative energy range. Thus, the weakening of the dependence of neutron surface absorption on neutron-proton asymmetry leads to a faster and stronger quenching of the $N =$ 20 gap as well as to a shift of the change of the $1f_{7/2} - 2p_{3/2}$ ordering to the region of higher $Z$ values. The break between the occupation probabilities of the last predominantly occupied $1d_{3/2}$ state and the first predominantly unoccupied state $2p_{3/2}$ also decreases in this case (see fig.~\ref{fig:Nnlj20}).

\begin{figure}
\resizebox{8,2 cm}{!}{%
\includegraphics{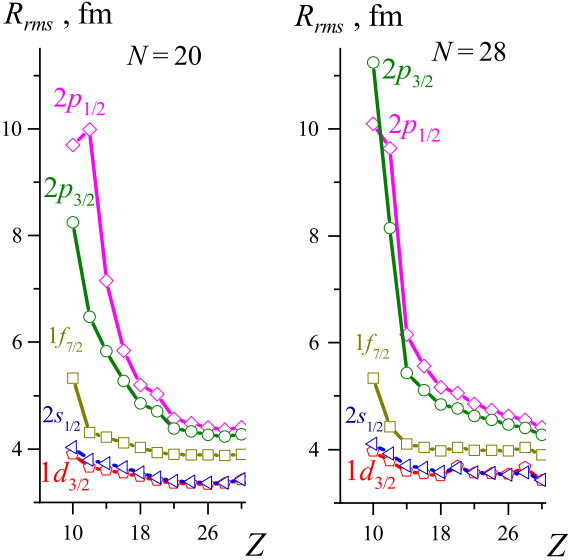}
}
\caption{Root mean-square radii $R_{rms}$ of orbits near the energy $E_F$ of the $N =$ 20, 28 isotones.}
\label{fig:Rrms}      
\end{figure}

The root mean-square radii $R_{rms}$ (\ref{Rrms}) of the orbits near the energy $E_F$, calculated with the global parameters KD, are shown in fig.~\ref{fig:Rrms} for the $N =$ 20, 28 isotones. The radius $R_{rms}$ of the $2p$ and $1f_{7/2}$ orbits increases sharply as $Z$ decreases. In particular, the $2p_{3/2}$ state of $^{32}$Mg (with the negative single-particle energy) is characterized by $R_{rms}$ almost 2 times larger than the $1d_{3/2}$ state. It should lead to the extension of the nuclear surface as it was proposed in \cite{Wang14, Ant05} for example. So, it is reliable to suggest that diffuseness $a_{HF}$ is greater than $a_{V}$ \cite{KD} in the range of nuclei with the sufficient neutron excess. We calculated the single-particle spectrum of $^{32}$Mg for different $a_{HF} = a_s$ values. The resulting gap between the $1d_{3/2}$ and $2p_{3/2}$ states of $^{32}$Mg decreased by 1.3 MeV when the diffuseness $a_{HF} = a_s$ increased from 0.67 to 0.8 fm.

\begin{figure}
\resizebox{8,2 cm}{!}{%
\includegraphics{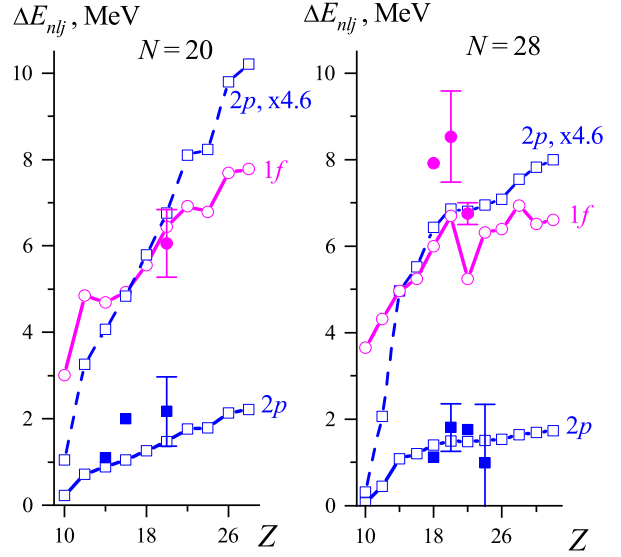}
}
\caption{Energy splitting of the $2p_{3/2} - 2p_{1/2}$ and $1f_{7/2} - 1f_{5/2}$ spin-orbit partners in isotones with $N =$ 20, 28. Solid symbols are the experimental data \cite{Bob10, Gau06, Bur14}, solid lines with open symbols are calculation with the DOMP, dashed line with open symbols are $2p$-splitting which has been multiplied by 4.6 for convenient comparison.}
\label{fig:Split}      
\end{figure}

The experimental data \cite{Gau06, Bur14}  indicate that the weakening of the splitting of the $2p_{3/2} - 2p_{1/2}$ spin-orbital partners is stronger than that of the $1f_{7/2} - 1f_{5/2}$ partners in $^{49}$Ca, $^{47}$Ar and $^{37}$S, $^{35}$Si isotones. In particular, no change in the $1f$-splitting was observed between $^{37}$S and $^{35}$Si isotones while change by 25\% in the $2p$-splitting was derived \cite{Bur14}. As key reasons, the influence of the features of two-body spin-orbit interaction in the case of $^{37}$S, $^{35}$Si isotones, as well as tensor forces and possible correlations in the case of $^{49}$Ca, $^{47}$Ar isotones was suggested. These experimental results were proposed \cite{Bur14} to be used for the testing of spin-orbital interaction of mean field theories. According to DOM, the calculated $2p$-splitting decreased more pronounced (in relative units) than the $1f$-splitting (see fig.~\ref{fig:Split}) for both $N =$ 20 and $N =$ 28 isotones. In the latter case, the filling of proton subshells has a significant effect on the neutron $1f$-splitting. Near the magic numbers $Z =$ 20, 28, the energy range near $E_F$, where the imaginary part of DOMP is equaled to 0, widens. As a result, neutron states below and above $E_F$ become more bound and less bound accordingly. Since the $1f_{7/2}$ state lies below $E_F$ and the $1f_{5/2}$ state lies above $E_F$ in the $N =$ 28 isotones, the calculated $2p$- and $1f$-splittings decreased almost identically from $^{48}$Ca to $^{46}$Ar. Note, that the calculations with the global parameters of DOM describe $Z$-dependence of the $1f$-, $2p$ -splitting correctly but underestimate the $1f$-splitting for $N =$ 28 near $Z =$ 20. To make quantitative agreement better, it’s necessary to individualize the DOMP parameters near the magic number $Z =$ 20. In contrast to the $N =$ 28 isotones, at $N =$ 20, $1f$ spin-orbit partners are both located above $E_F$ as well as $2p$ partners, and shell effects at $Z =$ 20, 28 influence on the $1f$- and $2p$-splittings in the same direction. Between $^{36}$S and $^{34}$Si, the calculated $2p$- and $1f$-splitting decreased by 16\% and 4.9\% respectively. So, the difference between them  is slightly less than the experimental data \cite{Bur14}. Note that at least qualitative agreement was obtained with the global parameters KD of the spin-orbit interaction.

\section{Summary }
\label{sec:4}
The evolution of the single-particle properties of the isotones with $N =$ 20, 28 was studied within the spherical DOM. Good agreement with the available experimental data for the stable nuclei was obtained with the global parameters of the imaginary and spin-orbit parts of the DOMP. These parameters  for the $N =$ 20, 28 isotones with the neutron excess lead to the following. The calculated $N =$ 20 and $N =$ 28 gaps reduce as $Z$ decreases, so that the $N =$ 28 gap completely disappears at $Z$ near 12, and then the change of the $1f_{7/2} - 2p_{3/2}$ ordering follows. As a result, the $1f_{7/2}$, $2p_{3/2}$ and $2p_{1/2}$ states practically degenerate in the $N =$ 28 isotones near the neutron drip line. The degeneracy of these states is one of the reasons for the deformation of the $N =$ 28 isotones due to the tensor-force-driven Jahn-Teller mechanism as it was shown earlier \cite{UO12}.

The change of the $1f_{7/2} - 2p_{3/2}$ ordering occurs inside of the neutron drip line in the $N =$ 20 isotones unlike the $N =$ 28 isotonic chain. The $N =$ 20 gap between the $1d_{3/2}$ and $2p_{3/2}$ states with $2j+1 = 4$   forms at $Z <$ 12. It can be related to the significant increase in the contribution of $4p4h$ excitations to the ground state of $^{32}$Mg \cite{Mac16, TS17} that determines its position on the shore of the island of inversion. So, the evolution of the single-particle energies in the spherical DOMP demonstrates the connection between the quenching of the $N =$ 28 shell closure and the appearance of the island of inversion at $N =$ 20. The faster decrease of $N =$ 20 shell gap was obtained as a result of the suggested weaker dependence (up to the independence) of the neutron surface absorption on neutron-proton asymmetry and increased diffuseness $a_{HF}$ at large neutron excess.

According to the DOM with the global parameters KD of the imaginary and spin-orbit potential, the $2p$-splitting reduces faster than the $1f$-splitting with $Z$ decrease. But this effect is less pronounced compared to the experimental data \cite{Bur14} between $^{36}$S and $^{34}$Si with $N =$ 20. For the $N =$ 28 isotones, the $Z =$ 20 and 28 shell closure significantly affects the value of the neutron $1f$-splitting.

The present study has demonstrated that DOM is a powerful tool to predict single-particle characteristics of nuclei far away from the $\beta$-stability region. The experimental data on the dependence of the surface absorption on neutron-proton asymmetry at large neutron (proton) excess is necessary for more accurate predictions of the single-particle spectra near the Fermi energy.

%
%
%

%
%

\end{document}